\documentclass[12pt,a4paper]{article}
\usepackage{graphicx}
\usepackage{amsmath}
\usepackage{amsfonts}
\usepackage{amssymb}
\newtheorem{theorem}{Theorem}

\newtheorem{conclusion}[theorem]{Conclusion}

\newtheorem{lemma}[theorem]{Lemma}

\newtheorem{proposition}[theorem]{Proposition}
\newtheorem{remark}[theorem]{Remark}

\newenvironment{proof}[1][Proof]{\textbf{#1.} }{\ \rule{0.5em}{0.5em}}

\begin{document}

\title{Quantum theory of human communication\\(PACS 89.65.-s)}
\date{}
\author{Wojtek Slowikowski\\Department of Mathematical Sciences University of Aarhus
\and Erik B. Nielsen\\Duborg School, Flensburg}
\maketitle
\begin{abstract}
We use notions and techniques of Quantum Field Theory to formulate and
investigate basic concepts and mechanisms of human communication. We start
with attitudes which correspond to photons frequencies, then we introduce
states-of-mind which correspond to wave functions. Finally, by way of the
second quantization, we come to states-of-opinions which correspond to states
of quantized radiation fields. In the present paper we shall only investigate
superpositions of pairs of coherent states (e.g. the government and the
opposition in a democratic country).
\end{abstract}
\tableofcontents

\section{\bigskip Introduction}

Suppose we want to investigate the social behavior of a human community
relative to a chosen subject. We expose the subject by producing a
questionnaire with entries containing descriptions of the components of the
subject in question. The questionnaire can be completed in several ways which
we shall call \emph{attitudes}. However, the process of filling in a
questionnaire in a specific way is not a deterministic procedure. Usually, the
mind of a respondent faces a number of preferences. Under casual influence
from outside the preferences can change. Hence, the reaction of a respondent
is not attached to a particular attitude but depends on the respondent's
\emph{state-of-mind}.

A statement of a respondent selecting an attitude or a state-of-mind shall
here be called a \emph{bit-of-information}.

Fix a finite set of frequences and consider photons with frequences from this
set. Hence the momentum space for these photons consists of finitely many
points and their wave-functions are superpositions of the frequencies. We
create our social model by way of the following substitutions,

1) bits-of-information replace photons

2) attitudes replace frequencies

3) states-of-mind replace wave-functions.

As we know, the instance of passing from frequencies to wave-functions, and
hence from attitudes to states-of-mind, constitutes the essence of the first quantization.

Totalities of states-of-mind produce new entities called \emph{opinions}. Then
\emph{states-of-opinion} emerge as the result of the process of the second
quantization which mathematically amounts to the repetition of the one used in
constructing a free quantum radiation field (cf.\cite{Berezin},
\cite{Louisell}). Hence

4) states-of-opinion replace quantized radiation fields.

To describe the second quantization we use the algebraic version of the
concept of the Bose-Fock space, the so-called Bose algebra (cf.\cite{Nielsen}%
). In \cite{Slowi} states-of-opinion were called information metabolisms.

The observables of the theory are just questions. Given a question and a
state-of-opinion, the occupation number formalism provides the expectation of
the number of positive answers to the question in a poll performed under the
given state-of-opinion (cf. \cite{Slowi-Nie}).

We treat bits-of-information circulating in human communities as bosons. The
propagators of those bits-of-information are the individual respondents which,
depending on their states-of-mind, provide answers yes or no to questions.
Also organizations can get the status of respondents and quality of possessing
a state-of-mind. Questions are coupled with orthogonal projections in the
space of states-of-mind. Affirmative answers are weighted by the assigned
number of energy-bits they carry: electing a Member of Parliament requires
many energy-bits in the form of single votes whereas a shareholder's single
vote carries the number of energy-bits equal to the number of owned shares.

In what follows we shall mainly be interested in superpositions of coherent
states-of-opinion. The coherent states considered in this paper are
mathematically identical with those of quantum optics. They are defined within
the polynomial representation of the Bose-Fock space (cf.\cite{Nielsen}). In
the present paper we restrict ourselves to analysis of states which are
superpositions of two coherent states, e.g. the government and the opposition
in a democratic country, the original inhabitants of a country and the
immigrants, Christians and Moslems etc. Such states will be called
\emph{bicoherent. }We shall show that the bicoherent states depend on two
parameters. The first, the \emph{interaction coefficient}, is a number between
$0$ and $1$ measuring the background for communication between respondents of
interacting fractions: a common language, traditions, religion, interest etc.
The second one is called the \emph{superposition constant}.

In a simple model constructed in Section \ref{model}, the superposition
constant plays a double role - if it is greater or equal to one, it prevents
interaction blocking the influence of the interaction coefficient. If
negative, it controls regions of high and low frequencies of affirmation to
questions asked in the superposition state-of-opinion. Moreover, under high
interaction a sudden critical switch of opinion can occur in consequence of a
minimal change of the superposition constant (cf. Remark \ref{switch}). We
also investigate change of opinion under temporary influence of some outside
factors (as for instance election campaigns). It is shown that such a
temporary influence diminishes high amplitudes (cf. Remark \ref{influence}).

The authors are much indebted to Krista Graversen for her help in editing this
paper. Also the first of the authors wishes to express his gratitude for the
hospitality of the Institute for Mathematical Behavioral Sciences at UCI.
Discussions with members of the Institute considerably influenced the form of
this paper. David Bulger pointed out some inaccuracies in our presentation of
the Bargmann Fock-space representation. We are grateful for his contribution.

\section{The first quantization}

\subsection{Profiles}

A selected sub-population characterized by a collection of \emph{attitudes}
will here be called a \emph{profile}. For example the body of parliament
members of a democratic country constitutes a profile. The attitudes will
represent different political affiliations. The states-of-mind will then
concern actual political problems. Also the government and its members can be
considered as a profile. Here the set of attitudes will include different
policies. For the profile of workers, the relevant attitudes will be concerned
with the unions.

Hence, the same physical population consists of several different profiles.
Profiles connected with a profession is easiest revealed by asking a question
to which the answer ''yes'' selects the states-of-mind of the profession. For
example, the question ''do you have a valid certificate qualifying you as a
physician?'' automatically extracts the profile of medical doctors. An
examination will filter respondents of the profile of a particular profession.
The whole population itself constitutes a profile as well.

\subsection{Attitudes}

Consider a community familiar with subjects which can be presented in a list
of statements. The statements can be accepted or rejected by members of the
community. In what follows we refer to this list of statements as a
\emph{questionnaire.} The term ''questionnaire'' should not be taken
literally. For instance, a questionnaire may consist of a set of examination
questions but it can as well be an ordinary questionnaire prepared for a poll.

A copy of a completed questionnaire shall be called an \emph{attitude}. The
quantum mechanical counterpart of an attitude is a frequency. Hence the
quantum mechanical equivalence of a space of attitudes is a momentum space
consisting of finitely many frequencies.

\subsection{The first quantization: from attitudes to states-of-mind}

Take a space of attitudes consisting of $n$ attitudes $\left\{
1,2,...,n\right\}  $. Consider the real-valued functions $x$ of $n$ real
variables $t_{1},t_{2},...,t_{n}$. To the attitude $j$ we attach the function
$e_{j},$ which is the value of the variable $t_{j},$
\[
e_{j}\left(  t_{1},t_{2,},...,t_{n}\right)  =t_{j}.
\]

We shall consider the real vector space $\mathcal{F}$ of vectors
\[
x=\lambda_{1}e_{1}+\lambda_{2}e_{2}+\cdots+\lambda_{k}e_{k},
\]
where $\lambda_{1},\lambda_{2},...,\lambda_{k}$ are arbitrary real numbers.
Given another vector from $\mathcal{F},$%
\[
y=\eta_{1}e_{1}+\eta_{2}e_{2}+\cdots+\eta_{k}e_{k},
\]
we define the inner product (Hermitian form) setting
\[
\left\langle x,y\right\rangle =\lambda_{1}\eta_{1}+\lambda_{2}\eta_{2}%
+\cdots+\lambda_{n}\eta_{n}%
\]
so that $e_{1},e_{2},...,e_{n}$ is an orthonormal basis in $\mathcal{F}$ and
each vector $x$ from $\mathcal{F}$ can be written in the form
\[
x=\left\langle x,e_{1}\right\rangle e_{1}+\left\langle x,e_{2}\right\rangle
e_{2}+\cdots+\left\langle x,e_{n}\right\rangle e_{n}.
\]
Then
\[
\left\langle x,y\right\rangle =\left\langle x,e_{1}\right\rangle \left\langle
y,e_{1}\right\rangle +\left\langle x,e_{2}\right\rangle \left\langle
y,e_{2}\right\rangle +\cdots+\left\langle x,e_{n}\right\rangle \left\langle
y,e_{n}\right\rangle .
\]
We shall write $\left|  x\right|  $ for the length of the vector $x$,
\[
\left|  x\right|  =\left\langle x,x\right\rangle ^{\frac{1}{2}}.
\]

A vector $x$ is called a \emph{state-of-mind} if $\left|  x\right|  =1.$ We do
not distinguish between states provided by $x$ and $-x$. Briefly we shall
write
\[
x_{/}\overset{def}{=}\frac{x}{\left|  x\right|  }%
\]
for the state-of-mind corresponding to the vector $x.$

If a respondent is in the state-of-mind $x,$ his attitude will be $j$ with
probability $\left\langle x,e_{j}\right\rangle ^{2},$ i.e. to the question
''which is your attitude?'' he will name the attitude $j$ with probability
$\left\langle x,e_{j}\right\rangle ^{2}.$

The (real) vector space $\mathcal{F}$ shall be called the space of
states-of-mind (as yet we have no interpretation for the process of
multiplication by the imaginary unit $i$).

Continuing the analogy with photons, the space $\mathcal{F}$ of states-of-mind
is the counterpart of the space of wave-functions (the state-space for
photons) depending on fixed finite number of frequencies. We just substitute
attitudes for frequencies.

Given states-of-mind $x$ and $y,$ the number $\left\langle x,y\right\rangle
^{2}$ is called the \emph{correlation} of $x$ and $y$\textbf{.} States for
which the correlation is equal to zero shall be called \emph{uncorrelated}.

Observe that the space $\mathcal{F}$ can be considered as the space of all
real-valued functions $x$ on the set $\left\{  1,2,...,n\right\}  ,$ each such
function assigning a real number $\lambda_{j}$ to $j$ from the set $\left\{
1,2,...,n\right\}  .$

\subsection{Questions as observables}

The process of assigning an attitude $j$ to a respondent can be ''first
quantized'' to a question directed to a respondent, ''are you fully accepting
the attitude $j$? '' The question itself then becomes an observable taking the
form of the projection
\[
Q_{e_{j}}=\left\langle e_{j},\cdot\right\rangle e_{j}.
\]
Now the procedure can be extended over arbitrary states-of-mind by attaching
to a state-of-mind $x$ the projection
\[
Q_{x}=\left\langle x,\cdot\right\rangle x
\]
which directed to a respondent runs as follows

\begin{center}
$"$are you in the state-of-mind $x$ ?''
\end{center}

We attach statistics to this question by way of the statement%

\[
\left\langle Q_{x}y,y\right\rangle =\left\langle x,y\right\rangle
^{2}=\left\{
\begin{array}
[c]{l}%
\text{the probability of obtaining the }\\
\text{answer ''yes'' to the question }Q_{x}\text{ }\\
\text{from a respondent in state }y
\end{array}
\right.
\]
i.e. the probability of the answer ''yes'' is equal to the correlation of $x$
and $y.$

More general questions $Q$ are linear combinations of questions of the form
$Q_{x}.$ Then
\[
\left\langle Qy,y\right\rangle =\left\{
\begin{array}
[c]{l}%
\text{the probability of obtaining the }\\
\text{answer ''yes'' to the question }Q\\
\text{from a respondent in state }y
\end{array}
\right.  .
\]
As an example we consider the projection
\[
Qx=\left\langle e_{i},x\right\rangle e_{i}+\left\langle e_{j},x\right\rangle
e_{\frak{j}},
\]
where $i$ and $j$ are different attitudes$\frak{.}$ The question corresponding
to this projection should read ''\textit{do you favor precisely the attitudes}
$i$ \textit{and} $j$ \textit{out of the collection of all possible
attitudes?}'' Here we have $Qx=x,$ exactly for $x=\left\langle e_{i}%
,x\right\rangle e_{i}+\left\langle e_{j},x\right\rangle e_{j}$ which means
that the answer ''yes'' comes with probability one from the states $x=\lambda
e_{i}+\eta e_{j},$ with $\lambda^{2}+\eta^{2}=1$.

As explained in the Introduction, each affirmative answer to a question
carries a number of energy-bits depending on the nature of the corresponding model.

\section{The second quantization}

The notions of attitude and state-of-mind concern individual respondents. The
second quantization provides a formalism by use of which the parallel notions
on the level of profiles can be defined (cf.\cite{Berezin}, \cite{Nielsen}).
The counterpart of the notion of attitude attached to an individual member of
a community will be the notion of opinion attached to a group of individuals.
Similarly the counterpart of the notion of state-of-mind attached to a
respondent will be the notion of state-of-opinion attached to a profile (which
can as well be the whole community). As a state-of-mind assigns a number to
every possible attitude, a state-of-opinion will assign a number to every
possible opinion of a profile, i.e. the states-of-opinion are functions over
the space of opinions. Given such a function, the square of its value on an
opinion gives the probability that the profile shares this opinion.

\subsection{Opinions and the Bose-Fock space for states-of-opinion}

Suppose that from a poll we have gathered information about the actual
distribution of attitudes of a profile. It means that we have a collection of
attitudes, where the same attitude may appear many times, a single time or not
at all. To obtain the precise definition we proceed as follows.

Let $\left\{  1,2,...,n\right\}  $ be the set of all attitudes. Then a tuple
of positive integers $\left(  k_{1},k_{2},...,k_{n}\right)  $ shall be called
an \emph{opinion} in which the attitude $j$ appears $k_{j}$ times for
$j=1,2,...,n$. If a particular attitude, say $i,$ does not appear at all, we
write $k_{i}=0.$ A poll assigns to each attitude the number of respondents
sharing this attitude i.e. it provides the opinion of the community. We use
the (real) Bargmann version of the Bose-Fock space construction. Write
$\widetilde{\mathcal{F}}$ for the algebra of all formal series
\[
f=\sum\lambda_{k_{1},k_{2},...,k_{n}}e_{k_{1},k_{2},...,k_{n}},
\]
where $e_{k_{1},k_{2},...,k_{n}}$ are products of variables $t_{1}%
,t_{2},,t_{n}:$
\[
e_{k_{1},k_{2},...,k_{n}}\left(  t_{1},t_{2},,t_{n}\right)  \overset{def}%
{=}t_{1}^{k_{1}}t_{2}^{k_{2}}\cdots t_{n}^{k_{n}}%
\]
and the sum runs through all the tuples $\left(  k_{1},k_{2},...,k_{n}\right)
$ of non-negative integers.

We multiply the series in the standard way setting
\[
e_{j_{1},j_{2},...,j_{n}}e_{k_{1},k_{2},...,k_{n}}=e_{j_{1}+k_{1},j_{2}%
+k_{2},...,j_{n}+k_{n},}.
\]
We postulate that a profile which consists of $k_{1}$ members carrying the
state-of-mind $e_{1},$ $k_{2}$ members carrying the state-of-mind $e_{2}$ etc.
up to $k_{n}$ members carrying the state-of-mind $e_{n},$ is in the
state-of-opinion $\frac{e_{k_{1},k_{2},...,k_{n}}}{\sqrt{k_{1}!k_{2}%
!...,k_{n}!}}.$

The set $\left\{  \frac{e_{k_{1},k_{2},...,k_{n}}}{\sqrt{k_{1}!k_{2}%
!...,k_{n}!}}\right\}  ,$ where $\left(  k_{1},k_{2},...,k_{n}\right)  $ runs
through all possible different opinions$, $ constitutes an orthonormal basis
in the Bargmann Bose-Fock space representation
\[
\Gamma\mathcal{F=}\left\{  f\in\widetilde{\text{ }\mathcal{F}}:\sum
_{k_{1},k_{2},...,k_{n}}\lambda_{k_{1},k_{2},...,k_{n}}^{2}k_{1}%
!k_{2}!...,k_{n}!<\infty\right\}  ,
\]
where
\[
\left\langle e_{j_{1},j_{2},...,j_{n}},e_{k_{1},k_{2},...,k_{n}}\right\rangle
=k_{1}!k_{2}!...,k_{n}!\delta_{j_{1},k_{1}}\delta_{j_{2},k_{2}}\cdots
\delta_{j_{n},k_{n}}.
\]
Hence, given
\begin{align*}
\sum_{j_{1},j_{2},...,j_{n}}\eta_{j_{1},j_{2},...,j_{n}}e_{j_{1}%
,j_{2},...,j_{n}}  &  \in\Gamma\mathcal{F}\\
\sum_{k_{1},k_{2},...,k_{n}}\lambda_{k_{1},k_{2},...,k_{n}}e_{k_{1}%
,k_{2},...,k_{n}}  &  \in\Gamma\mathcal{F},
\end{align*}
$\ $\ we have
\begin{align*}
&  \left\langle \sum_{j_{1},j_{2},...,j_{n}}\eta_{j_{1},j_{2},...,j_{n}%
}e_{j_{1},j_{2},...,j_{n}},\sum_{k_{1},k_{2},...,k_{n}}\lambda_{k_{1}%
,k_{2},...,k_{n}}e_{k_{1},k_{2},...,k_{n}}\right\rangle \\
&  =\sum_{j_{1},j_{2},...,j_{n}}\sum_{k_{1},k_{2},...,k_{n}}\eta_{j_{1}%
,j_{2},...,j_{n}}\lambda_{k_{1},k_{2},...,k_{n}}\left\langle e_{j_{1}%
,j_{2},...,j_{n}},e_{k_{1},k_{2},...,kn}\right\rangle \\
&  =\sum_{j_{1},j_{2},...,j_{n}}\sum_{k_{1},k_{2},...,k_{n}}\eta_{j_{1}%
,j_{2},...,j_{n}}\lambda_{k_{1},k_{2},...,k_{n}}k_{1}!k_{2}!...,k_{n}%
!\delta_{j_{1},k_{1}}\delta_{j_{2},k_{2}}\cdots\delta_{j_{n},k_{n}}\\
&  =\sum_{k_{1},k_{2},...,k_{n}}\eta_{k_{1},k_{2},...,k_{n}}\lambda
_{k_{1},k_{2},...,k_{n}}k_{1}!k_{2}!...,k_{n}!.
\end{align*}
$\ \ $For $x_{1},x_{2},\cdots,x_{n},y$ from $\mathcal{F}$ we have the
following useful formula (cf.\cite{Nielsen})
\[
\left\langle x_{1}x_{2}\cdots x_{n},y^{m}\right\rangle =\left\{
\begin{array}
[c]{cc}%
m!\left\langle x_{1},y\right\rangle \left\langle x_{2},y\right\rangle
\cdots\left\langle x_{m},y\right\rangle  & \text{for }m=n\\
0 & \text{otherwise}%
\end{array}
\right.  .
\]

A \emph{state-of-opinion} will be a vector $f$ from $\Gamma\mathcal{F}$ such
that $\left\langle f,f\right\rangle =1.$ This way, for a state-of-opinion
$f=\sum$ $\lambda_{k_{1},k_{2},...,k_{n}}e_{k_{1},k_{2},...,k_{n}}$ we have
$\sum\lambda_{k_{1},k_{2},...,k_{n}}^{2}=1,$ and for each opinion $\left(
k_{1},k_{2},...,k_{n}\right)  $ the number $\lambda_{k_{1},k_{2},...,k_{n}%
}^{2}$ represents the probability that the members of the concerned profile
share the opinion $\left(  k_{1},k_{2},...,k_{n}\right)  .$ We identify
states-of-opinion $f$ and $-f.$ If all $k_{j}=0,$ then we get the vector
$\phi,$ $\;\phi\left(  t_{1},t_{2},...,t_{n}\right)  =1,$ called the
\emph{vacuum vector}.

Notice that states-of-opinion can be interpreted as functions defined on the
space of opinions, each such function assigning to an opinion $\left(
k_{1},k_{2},...,k_{n}\right)  $ a real number $\lambda_{k_{1},k_{2},...,k_{n}}.$

\begin{remark}
In the present paper there is no need to take for $\lambda_{k_{1}%
,k_{2},...,k_{n}}$ the complex numbers. Should such a need occur in the
future, the necessary adjustments are elementary.
\end{remark}

We shall need the notion of the operator $w^{\ast}$of $\emph{annihilation}$ by
an element $w$ from $\mathcal{F}$ (cf.\cite{Nielsen})$.$ We define $w^{\ast}$
first for the basis vectors $e_{j}$ of $\mathcal{F}$ setting for
$f\in\mathcal{\Gamma F}$
\[
\left(  e_{j}^{\ast}f\right)  \left(  t_{1},t_{2},...,t_{n}\right)
=\frac{\partial}{\partial t_{j}}f\left(  t_{1},t_{2},...,t_{n}\right)  ,
\]
and then extend it linearly to include all $w$ from $\mathcal{F}.$

The only infinite sums we will use are the elements of $\Gamma\mathcal{F}$
called \emph{coherent vectors}, which are the exponential functions
\[
e^{x}=\sum_{n=0}^{\infty}\frac{1}{n!}x^{n}%
\]
of $x\in\mathcal{F}$. It is easy to verify that
\[
\left\langle e^{x},e^{y}\right\rangle =e^{\left\langle x,y\right\rangle }.
\]

\subsection{Occupation numbers and their statistics}

To every orthogonal projection $Q$ in $\mathcal{F}$ and every natural number
$k$ we assign a projection $Q^{\left(  k\right)  }$ in $\Gamma\mathcal{F}$
which we define as follows:

Take $x_{1},...,x_{p},y_{1},...,y_{q}\in\mathcal{F}$ such that $Qx_{j}=x_{j} $
for $j=1,2,...,p$ and $Qy_{i}=0$ for $i=1,2,...,q.$ Then we define%

\[
Q^{\left(  k\right)  }\left(  x_{1}x_{2}\cdots x_{p}y_{1}y_{2}\cdots
y_{q}\right)  \overset{def}{=}\left\{
\begin{array}
[c]{cc}%
x_{1}x_{2}\cdots x_{p}y_{1}y_{2}\cdots y_{q} & p=k\\
0 & \text{otherwise}%
\end{array}
\right.  .
\]
It is easy to extend $Q^{\left(  k\right)  }$ to an orthogonal projection in
$\Gamma\mathcal{F}$. The projection $Q^{\left(  k\right)  }$ is an observable
in the space of states-of-opinion and corresponds to the question:%

\[
\text{Is there exactly }k\text{ answers ''yes'' to the question }Q\text{?}%
\]
Consequently, for a state-of-opinion $f$ we have%

\[
\left\langle Q^{\left(  k\right)  }f,f\right\rangle =\left\{
\begin{tabular}
[c]{l}%
the probability that in the state $f$\\
we get precisely $k$ answers ''yes'' to $Q$%
\end{tabular}
\right.  .
\]
Let $f$ be a state, i.e. let $\left|  f\right|  =1.$ The numbers $\left\langle
Q^{\left(  k\right)  }f,f\right\rangle $ are called the \emph{occupation
numbers} of affirmation of $Q$ in the state $f.$

We extend $Q$ to a derivation $d\Gamma Q,$ i.e. a transformation obeying the
Leibniz rule,
\[
\left(  d\Gamma Q\right)  fg=\left(  d\Gamma Qf\right)  g+f\left(  d\Gamma
Qg\right)  .
\]
This operation is often called the \emph{second quantization} of $Q.$ It is
easy to verify that the spectral decomposition of $d\Gamma Q$ is
\[
d\Gamma Q=\Sigma_{k=0}^{\infty}kQ^{\left(  k\right)  }.
\]
Hence, if $f$ is a state-of-opinion, then
\begin{equation}
\left\langle d\Gamma Qf,f\right\rangle =\Sigma_{k=0}^{\infty}k\left\langle
Q^{\left(  k\right)  }f,f\right\rangle =\left\{
\begin{tabular}
[c]{l}%
the expected number of energy-bits\\
coming from the affirmative\\
answers to $Q$ in the state-of-opinion $f.$%
\end{tabular}
\right.  \label{expect}%
\end{equation}
However, it is not the expected number of energy-bits coming from the
affirmative answers which is measured by a poll but the expected percentage
$\mathcal{R}\left(  Q,f\right)  $ of those energy-bits,
\begin{equation}
\mathcal{R}\left(  Q,f\right)  =\frac{\left\langle d\Gamma Qf,f\right\rangle
}{\left\langle d\Gamma If,f\right\rangle }=\left\{
\begin{tabular}
[c]{l}%
the expected percentage of energy-\\
bits coming from the affirmative\\
answers to $Q$ in the state-of-opinion $f.$%
\end{tabular}
\right.  \label{rel-expc}%
\end{equation}
We shall call $\mathcal{R}\left(  Q,f\right)  $ the \emph{relative expectation
}for the energy of affirmation of\emph{\ }$Q$ in the state $f.$ Here the
identity operator $I$ corresponds to the question: ''How many energy-bits are available''?

Given a state-of-opinion $f,$ we can produce a new one by making a
superposition of $f$ with the vacuum
\[
\left(  f+\alpha\text{$\phi$}\right)  _{/}=\frac{f+\alpha\text{$\phi$}}%
{\sqrt{1+\alpha^{2}+2\left\langle \text{$\phi$},f\right\rangle }}.
\]
Since
\[
\left\langle d\Gamma Q\left(  f+\alpha\text{$\phi$}\right)  ,f+\alpha
\text{$\phi$}\right\rangle =\left\langle d\Gamma Qf,f\right\rangle ,
\]
we get
\[
\mathcal{R}\left(  Q,\left(  f+\alpha\text{$\phi$}\right)  _{/}\right)
=\frac{\left\langle d\Gamma Q\left(  f+\alpha\text{$\phi$}\right)
_{/},\left(  f+\alpha\text{$\phi$}\right)  _{/}\right\rangle }{\left\langle
d\Gamma I\left(  f+\alpha\text{$\phi$}\right)  _{/},\left(  f+\alpha
\text{$\phi$}\right)  _{/}\right\rangle }=\frac{\left\langle d\Gamma
Qf,f\right\rangle }{\left\langle d\Gamma If,f\right\rangle }=\mathcal{R}%
\left(  Q,f\right)
\]
which means that the superposition with the vacuum does not change the
percentage of energy-bits coming from affirmation of $Q.$

\subsection{Coherent states}

A coherent state-of-opinion describes respondents with states-of-mind
concentrated around a special state-of-mind called the mode of coherence, e.g.
physicians with their professional curriculum as the mode, members of a
political party with their party program as the mode, lawyers with their
professional know-how as the mode etc.

Take a vector $x$ from the states-of-mind space $\mathcal{F}.$ The
\emph{coherent state} $c\left(  x\right)  $ generated by $x$ is the normalized
coherent vector $e^{x},$
\begin{equation}%
\begin{array}
[c]{l}%
c\left(  x\right)  =e_{/}^{x}=e^{-\frac{1}{2}\left\langle x,x\right\rangle
}e^{x}\\
c\left(  0\right)  =\text{$\phi$}.
\end{array}
\nonumber
\end{equation}
Observe that if the number $\left\langle x-y,x-y\right\rangle $ is very large,
the correlation
\begin{equation}
\left\langle c\left(  x\right)  ,c\left(  y\right)  \right\rangle
=e^{-\frac{1}{2}\left|  x-y\right|  ^{2}} \label{corr.c(x)-c(y)}%
\end{equation}
is almost $0,$ i.e. $c\left(  x\right)  $ and $c\left(  y\right)  $ are almost
uncorrelated. Hence any experiment performed in one of those states has almost
no probable relation to an experiment performed in the other state.

The coherent states are ''almost'' multiplicative; we have
\[
c\left(  x+y\right)  =e^{-\left\langle x,y\right\rangle }c\left(  x\right)
c\left(  y\right)  .
\]

The number $\left|  x\right|  ^{2},$ the state-of-mind $x_{/},$ and the vector
$x$ shall be respectively called the \emph{energy}, the \emph{mode} and the
\emph{generating vector\ }of the coherent state $c\left(  x\right)  .$ Hence
in the background of a given coherent state lies the mode which is the
state-of-mind that provides the right frequencies of occurrence of the
attitudes from a fixed list. The mode for a given coherent state can be
approximated as follows. We produce a ''super-questionnaire'' out of all
involved attitudes; then count the frequencies of the choice of particular
attitudes in a poll and take their square roots as coefficients to$\ $the
respective attitude.

We can easily compute the relative expectation $\mathcal{R}$ for $Q$ in a
coherent state $c\left(  x\right)  .$ Since $d\Gamma Q$ is a derivation, we have%

\[
d\Gamma Qc\left(  x\right)  =\left(  Qx\right)  c\left(  x\right)
\]
so that
\[
\left\langle d\Gamma Qc\left(  x\right)  ,c\left(  x\right)  \right\rangle
=\left\langle x,Qx\right\rangle =\left|  Qx\right|  ^{2}%
\]
and we obtain the number
\[
\mathcal{R}\left(  Q,c\left(  x\right)  \right)  =\left\langle x_{/}%
,Qx_{/}\right\rangle =\left|  Qx_{/}\right|  ^{2}%
\]
which does not depend on the energy $\left|  x\right|  ^{2}$ of $c\left(
x\right)  .$

\section{Bicoherence}

The concept of bicoherence concerns a community consisting of two coherent
fractions, e.g. the government and the opposition in a democratic country,
members of two different religious affiliations, a population consisting of
natives and immigrants etc. In each of these cases the state-of-opinion of the
whole population is a superposition of the states-of-opinion of two coherent
sub-profiles. The state-of-opinion of the superposition is not any longer
coherent and shall be called bicoherent.

One can easily quote important cases involving more than two coherent states
but already in the case of three, the amount of necessary computation will
double the size of this paper and hence must be postponed to a separate publication.

\subsection{Bicoherent states}

Take coherent states $c\left(  u\right)  $ and $c\left(  v\right)  ,$ $u\neq
v,$ and a number $\lambda.$ The number
\[
\omega=\left\langle c\left(  u\right)  ,c\left(  v\right)  \right\rangle
=e^{-\frac{1}{2}\left|  u-v\right|  ^{2}}%
\]
shall be called the \emph{interaction coefficient}. States of the form
\begin{equation}
c_{\lambda}\left(  u,v\right)  =\frac{c\left(  u\right)  +\lambda c\left(
v\right)  }{\vartheta\left(  \lambda,\omega\right)  },
\label{bicoh-state-formula}%
\end{equation}
where
\begin{equation}
\vartheta\left(  \lambda,\omega\right)  =\left|  c\left(  u\right)  +\lambda
c\left(  v\right)  \right|  =\sqrt{1+\lambda^{2}+2\lambda\omega},
\label{theta}%
\end{equation}
shall be called \emph{bicoherent states}. For $\lambda\neq0$ we have
\[
c_{\lambda}\left(  u,v\right)  =c_{\frac{1}{\lambda}}\left(  v,u\right)
\]
so that for $\lambda$ close to infinity, $c_{\lambda}\left(  u,v\right)  $
behaves exactly as $c_{\lambda}\left(  v,u\right)  $ behaves for $\lambda$
close to zero. The coefficient $\lambda$ will be called the
\emph{superposition constant}.

The closer to zero is $\omega,$ i.e. the greater is $\left|  u-v\right|  ,$
the more the states $c\left(  u\right)  $ and $c\left(  v\right)  $ act as
uncorrelated, and $c_{\lambda}\left(  u,v\right)  $ describes a profile split
into two groups which hardly communicate with each other.

With fixed $u$ and $v$, when $\lambda$ increases to infinity, the state
$c_{\lambda}\left(  u,v\right)  $ converges to the state $c\left(  v\right)
,$ and when $\lambda$ decreases to zero, it converges to the state $c\left(
u\right)  .$ Excluding the case of simultaneous $\lambda=-1$ and $u=v,$ we get
from \ref{expect} the expected number of energy-bits of the affirmative
answers to a question $Q:$
\begin{align*}
&  \ \left\langle c_{\lambda}\left(  u,v\right)  ,\left(  d\Gamma Q\right)
c_{\lambda}\left(  u,v\right)  \right\rangle \\
&  =\frac{\kappa\left(  Q;\lambda,u,v,\omega\right)  }{\vartheta\left(
\lambda,\omega\right)  ^{2}},
\end{align*}
where
\begin{equation}
\kappa\left(  Q;\lambda,u,v,\omega\right)  =\left|  Qu\right|  ^{2}%
+2\lambda\omega\left\langle Qu,v\right\rangle +\lambda^{2}\left|  Qv\right|
^{2}. \label{kappa}%
\end{equation}
Applying (\ref{rel-expc}) we get
\begin{align}
&  \mathcal{R}\left(  Q,c_{_{\lambda}}\left(  u,v\right)  \right) \nonumber\\
&  =\left\{
\begin{tabular}
[c]{l}%
the $\text{expected percentage of affirmations }$\\
of $Q\text{ in }$the state-of-opinion $c_{\lambda}\left(  u,v\right)  $%
\end{tabular}
\right. \label{dG-.-}\\
&  =\frac{\kappa\left(  Q;\lambda,u,v,\omega\right)  }{\kappa\left(
I;\lambda,u,v,\omega\right)  }.\nonumber
\end{align}
Define
\begin{align}
\mathcal{R}_{\omega}\left(  Q,\lambda,u,v\right)   &  =\frac{\left|
Qu\right|  ^{2}+\lambda^{2}\left|  Qv\right|  ^{2}+2\lambda\left\langle
Qu,v\right\rangle \omega}{\left|  u\right|  ^{2}+\lambda^{2}\left|  v\right|
^{2}+2\lambda\left\langle u,v\right\rangle \omega}\label{R-sub-omega}\\
&  =\frac{\left|  Qu_{/}\right|  ^{2}+\lambda^{2}\left|  Qv_{/}\right|
^{2}+2\lambda\left\langle Qu_{/},v_{/}\right\rangle \omega}{1+\lambda
^{2}+2\lambda\left\langle u_{/},v_{/}\right\rangle \omega}.\nonumber
\end{align}
\emph{\ }Then choosing $\omega=e^{-\frac{1}{2}t^{2}\left|  u-v\right|  ^{2}}$
we get
\begin{align}
&  \mathcal{R}_{\omega}\left(  Q,\lambda,u,v\right) \label{omega-no-omega}\\
&  =\frac{\left|  Qu\right|  ^{2}+\lambda^{2}\left|  Qv\right|  ^{2}%
+2\lambda\left\langle Qu,v\right\rangle e^{-\frac{1}{2}t^{2}\left|
u-v\right|  ^{2}}}{\left|  u\right|  ^{2}+\lambda^{2}\left|  v\right|
^{2}+2\lambda\left\langle u,v\right\rangle e^{-\frac{1}{2}t^{2}\left|
u-v\right|  ^{2}}}=\mathcal{R}\left(  Q,c_{_{\lambda}}\left(  tu,tv\right)
\right)  ,\nonumber
\end{align}
where $\omega\ $from the open interval $\left(  0,1\right)  \ $can now be
treated as an independent variable modulo an adjustment of the amplitude of
$u-v$.

The interaction coefficient measures the ability for interaction (as for
instance speaking the same everyday language, being a citizen of a democratic
country, having the same cultural or religious background etc.).

The superposition constant plays two different roles. It measures the degrees
of influence the participating coherent states have on their superposition.
And it marks the existence of wish to enter the interaction at all: Catholics
and Protestants of Northern Ireland are fully capable of interacting on an
arbitrarily high social level but they will not enter the interaction due to
some special reasons.

Suppose that some social forces alter the coherent state $c\left(  x\right)  $
to another coherent state $c\left(  y\right)  .$ Then, writing $z=y-x,$ we can
consider $z$ as the vector altering the generating vector $x$ of the given
coherent state to a new generating vector $x+z$ of the new coherent state
$c\left(  x+z\right)  =c\left(  y\right)  .$ This reduces the process of
changing $c\left(  x\right)  $ into $c\left(  y\right)  $ to the application
of the transformation $W_{z}$ dependent on a vector $z$ from $\mathcal{F}%
\mathbf{.}$ The transformation
\[
W_{z}c\left(  x\right)  =c\left(  x+z\right)
\]
of $c\left(  x\right)  $ into $c\left(  x+z\right)  $ is called the \emph{Weyl
transformation. }Given $z,$ the Weyl transformation $W_{z}$ is uniquely
extendable to a linear isometry (states-of-mind preserving transformation) of
$\Gamma\mathcal{F}$ onto itself (cf. \cite{Nielsen}). The Weyl transformation
$W_{z}$ is fully described by the coherent state $c\left(  z\right)  $ which
shall be called the \emph{generator }of $W_{z}.$

\subsection{The mathematics of bicoherence}

In this section we shall prove a series of results necessary for further
development of the theory.

Let $y,u\in\mathcal{F}$ and $f,g\in\Gamma\mathcal{F}$ and let $Q$ be an
orthogonal projection. In the proofs below we shall freely use the following
identities (cf.\cite{Nielsen}):%

\begin{align*}
\left\langle yf,g\right\rangle  &  =\left\langle f,y^{\ast}g\right\rangle \\
y^{\ast}\left(  fg\right)   &  =\left(  y^{\ast}f\right)  g+f\left(  y^{\ast
}g\right) \\
y^{\ast}c\left(  u\right)   &  =\left\langle y,u\right\rangle c\left(
u\right) \\
\left\langle d\Gamma Qf,g\right\rangle  &  =\left\langle f,d\Gamma
Qg\right\rangle \\
d\Gamma Q\left(  fg\right)   &  =\left(  d\Gamma Qf\right)  g+f\left(  d\Gamma
Qg\right) \\
d\Gamma Qc\left(  u\right)   &  =\left(  Qu\right)  c\left(  u\right)  .
\end{align*}

Given $x,z\in\mathcal{F},$ $\left|  z\right|  =1,$ we briefly write
\[
z\#c\left(  x\right)  =\left(  \left\langle x,z\right\rangle \text{$\phi$%
}-z\right)  c\left(  x\right)  .
\]

\begin{lemma}
Take $u,z\in\mathcal{F}.$ If $\left|  z\right|  =1$,$\ $then the vector
$z\#c\left(  u\right)  $ is a state-of-opinion.
\end{lemma}%

\proof
We have
\[
\left\langle c\left(  u\right)  ,zc\left(  u\right)  \right\rangle
=\left\langle z^{\ast}c\left(  u\right)  ,c\left(  u\right)  \right\rangle
=\left\langle \left\langle z,u\right\rangle c\left(  u\right)  ,c\left(
u\right)  \right\rangle =\left\langle z,u\right\rangle
\]
and
\begin{align*}
&  \left\langle zc\left(  u\right)  ,zc\left(  u\right)  \right\rangle \\
&  =\left\langle c\left(  u\right)  ,z^{\ast}\left(  zc\left(  u\right)
\right)  \right\rangle =\left|  z\right|  ^{2}+\left\langle z,u\right\rangle
\left\langle c\left(  u\right)  ,zc\left(  u\right)  \right\rangle =\left|
z\right|  ^{2}+\left\langle z,u\right\rangle ^{2}%
\end{align*}
so that
\begin{align*}
&  \left\langle \left(  \left\langle u,z\right\rangle \text{$\phi$}-z\right)
c\left(  u\right)  ,\left(  \left\langle u,z\right\rangle \text{$\phi$%
}-z\right)  c\left(  u\right)  \right\rangle \\
&  =\left\langle \left\langle u,z\right\rangle c\left(  u\right)  -zc\left(
u\right)  ,\left\langle u,z\right\rangle c\left(  u\right)  -zc\left(
u\right)  \right\rangle \\
&  =\left\langle u,z\right\rangle ^{2}-2\left\langle u,z\right\rangle
\left\langle c\left(  u\right)  ,zc\left(  u\right)  \right\rangle
+\left\langle zc\left(  u\right)  ,zc\left(  u\right)  \right\rangle \\
&  =\left\langle u,z\right\rangle ^{2}-2\left\langle u,z\right\rangle
^{2}+\left|  z\right|  ^{2}+\left\langle z,u\right\rangle ^{2}=\left|
z\right|  ^{2}.
\end{align*}%

\endproof

\bigskip

Now we can verify the following

\begin{proposition}
We have
\[
\underset{\alpha\rightarrow0}{\lim}\left|  \frac{c\left(  u+\alpha z\right)
-c\left(  u\right)  }{\sqrt{2}\sqrt{1-e^{-\frac{1}{2}\left(  \alpha\left|
z\right|  \right)  ^{2}}}}-z\#c\left(  u\right)  \right|  =0
\]
i.e. the bicoherent states $c_{-1}\left(  u+\alpha z,u\right)  $ converge
strongly to the state $z\#c\left(  u\right)  .$
\end{proposition}%

\proof
Take an arbitrary $y\in\mathcal{F}.$ Using l'Hospital Theorem, we get
\[
\underset{\alpha\rightarrow0}{\lim}\left\langle \frac{c\left(  u+\alpha
z\right)  -c\left(  u\right)  }{\sqrt{2}\sqrt{1-e^{-\frac{1}{2}\left(
\alpha\left|  z\right|  \right)  ^{2}}}}-\left(  \left\langle u,z_{/}%
\right\rangle \text{$\phi$}-z_{/}\right)  c\left(  u\right)  ,c\left(
y\right)  \right\rangle =0.
\]
But $z_{/}\#c\left(  u\right)  $ lies on the unit sphere and $\left\{
e^{x}:x\in\mathcal{F}\right\}  $ is total so that the Proposition holds.%
\endproof

\begin{proposition}
We have
\[
\ \underset{\alpha\rightarrow0}{\lim}\left\langle c_{-1}\left(  u+\alpha
z,u\right)  ,\left(  d\Gamma Q\right)  c_{-1}\left(  u+\alpha z,u\right)
\right\rangle =\left|  Qu\right|  ^{2}+\left|  Qz_{/}\right|  ^{2}.
\]
\end{proposition}%

\proof
Indeed,
\begin{align*}
&  \left\langle c\left(  u+\alpha z\right)  -c\left(  u\right)  ,d\Gamma
Q\left(  c\left(  u+\alpha z\right)  -c\left(  u\right)  \right)
\right\rangle \\
&  =\left\langle c\left(  u+\alpha z\right)  ,\left(  Q\left(  u+\alpha
z\right)  \right)  c\left(  u+\alpha z\right)  \right\rangle -\left\langle
c\left(  u+\alpha z\right)  ,\left(  Qu\right)  c\left(  u\right)
\right\rangle \\
&  -\left\langle c\left(  u\right)  ,\left(  Q\left(  u+\alpha z\right)
\right)  c\left(  u+\alpha z\right)  \right\rangle +\left\langle c\left(
u\right)  ,\left(  Qu\right)  c\left(  u\right)  \right\rangle \\
&  =\left\langle Q\left(  u+\alpha z\right)  ,u+\alpha z\right\rangle
-2\left\langle Qu,u+\alpha z\right\rangle e^{-\frac{1}{2}\alpha^{2}\left|
z\right|  ^{2}}+\left\langle Qu,u\right\rangle ,
\end{align*}
and using l'Hospital Theorem, we get
\[
\underset{\alpha\rightarrow0}{\lim}\frac{\left|  Q\left(  u+\alpha z\right)
\right|  ^{2}-2\left\langle Qu,u+\alpha z\right\rangle e^{-\frac{1}{2}%
\alpha^{2}\left|  z\right|  ^{2}}+\left|  Qu\right|  ^{2}}{2\left(
1-e^{-\frac{1}{2}\left(  \alpha\left|  z\right|  \right)  ^{2}}\right)
}=\left|  Qu\right|  ^{2}+\left|  Qz_{/}\right|  ^{2}.
\]%

\endproof

\begin{proposition}
\label{lim-state}We have
\[
\underset{\alpha\rightarrow0}{\lim}\left(  c_{\lambda}\left(  u,v\right)
-W_{\alpha z}c_{\lambda}\left(  u,v\right)  \right)  _{/}=\left(
z_{/}\#c\left(  u\right)  +\lambda z_{/}\#c\left(  v\right)  \right)  _{/}.
\]
\end{proposition}%

\proof
We have
\[
W_{\alpha z}c_{\lambda}\left(  u,v\right)  =\frac{c\left(  u+\alpha z\right)
+\lambda c\left(  v+\alpha z\right)  }{\vartheta\left(  \lambda,\omega\right)
},
\]
where $\vartheta$ is given by (\ref{theta}). Let
\begin{align*}
U_{\alpha}  &  =c\left(  u\right)  -c\left(  u+\alpha z\right) \\
V_{\alpha}  &  =c\left(  v\right)  -c\left(  v+\alpha z\right) \\
\left|  U_{\alpha}\right|  ^{2}  &  =2\left(  1-e^{-\frac{1}{2}\left|  \alpha
z\right|  ^{2}}\right)  =\left|  V_{\alpha}\right|  ^{2}.
\end{align*}
Then
\begin{align*}
\frac{U_{\alpha}}{\left|  U_{\alpha}\right|  }  &  \rightarrow U=\left(
\left\langle u,z_{/}\right\rangle \text{$\phi$}-z_{/}\right)  c\left(
u\right)  ,\;\\
\;\;\frac{V_{\alpha}}{\left|  V_{\alpha}\right|  }  &  \rightarrow V=\left(
\left\langle v,z_{/}\right\rangle \text{$\phi$}-z_{/}\right)  c\left(
v\right)
\end{align*}%
\begin{align*}
&  c_{\lambda}\left(  u,v\right)  -W_{\alpha z}c_{\lambda}\left(  u,v\right)
\\
&  =\frac{\left(  c\left(  u\right)  -c\left(  u+\alpha z\right)  \right)
+\lambda\left(  c\left(  v\right)  -c\left(  v+\alpha z\right)  \right)
}{\vartheta\left(  \lambda,\omega\right)  }\\
&  =\frac{U_{\alpha}+\lambda V_{\alpha}}{\vartheta\left(  \lambda
,\omega\right)  }%
\end{align*}%
\begin{align*}
&  \frac{c_{\lambda}\left(  u,v\right)  -W_{\alpha z}c_{\lambda}\left(
u,v\right)  }{\left|  c_{\lambda}\left(  u,v\right)  -W_{\alpha z}c_{\lambda
}\left(  u,v\right)  \right|  }\\
&  =\frac{U_{\alpha}+\lambda V_{\alpha}}{\left|  U_{\alpha}+\lambda V_{\alpha
}\right|  }=\frac{\frac{U_{\alpha}}{\left|  U_{\alpha}\right|  }+\lambda
\frac{V_{\alpha}}{\left|  V_{\alpha}\right|  }}{\left|  \frac{U_{\alpha}%
}{\left|  U_{\alpha}\right|  }+\lambda\frac{V_{\alpha}}{\left|  V_{\alpha
}\right|  }\right|  }\rightarrow\frac{U+\lambda V}{\left|  U+\lambda V\right|
}\\
&  =\frac{\left(  \left\langle u,z_{/}\right\rangle \text{$\phi$}%
-z_{/}\right)  c\left(  u\right)  +\lambda\left(  \left\langle v,z_{/}%
\right\rangle \text{$\phi$}-z_{/}\right)  c\left(  v\right)  }{\left|  \left(
\left\langle u,z_{/}\right\rangle \text{$\phi$}-z_{/}\right)  c\left(
u\right)  +\lambda\left(  \left\langle v,z_{/}\right\rangle \text{$\phi$%
}-z_{/}\right)  c\left(  v\right)  \right|  }.
\end{align*}%

\endproof

\bigskip

Given a real number $\lambda$ and $u,v,z\in\mathcal{F},$ we define
\[
\iota\left(  Q,u,v,z\right)  =\left\langle v-u,z\right\rangle \left(
\left\langle Qu,v\right\rangle \left\langle v-u,z\right\rangle +\left\langle
\left(  v-u\right)  ,Qz\right\rangle \right)  .
\]

Take $u,v,z\in\mathcal{F}$, where $\left|  z\right|  =1.$ Then
\begin{align}
&  \left\langle dQ\left(  z\#c\left(  u\right)  +\lambda z\#c\left(  v\right)
\right)  ,z\#c\left(  u\right)  +\lambda z\#c\left(  v\right)  \right\rangle
\label{dGafter}\\
&  =\vartheta^{2}\left(  \lambda,\omega\right)  \left|  Q\left(  z\right)
\right|  ^{2}+\kappa\left(  Q;\lambda,u,v,\omega\right)  -2\lambda\omega
\iota\left(  Q,u,v,z\right)  .\nonumber
\end{align}%

\proof
By direct computations we verify the relation
\begin{align*}
&  \frac{1}{\omega}\left\langle d\Gamma Qz\#c\left(  u\right)  ,z\#c\left(
v\right)  \right\rangle \\
&  =\left\langle z,Qz\right\rangle +\left\langle z,z\right\rangle \left\langle
Qu,v\right\rangle +\iota\left(  Q,u,v,z\right)  .
\end{align*}
which applied to the left side of (\ref{dGafter}) yields its right side.%
\endproof

\subsection{Consequences of temporary external influence}

Consider a profile in a bicoherent state $c_{\lambda}\left(  u,v\right)  $ and
an element $z\in\mathcal{F}.$ For $\alpha>0,$ let an external influence caused
by $W_{\alpha z}$ ,
\[
c_{\lambda}\left(  u,v\right)  \rightarrow W_{\alpha z}c_{\lambda}\left(
u,v\right)  ,
\]
induce a new state $c_{\lambda}\left(  u+\alpha z,v+\alpha z\right)  $. In
consequence of the enforcement, the population falls into the superposition
state
\[
\left(  c_{\lambda}\left(  u+\alpha z,v+\alpha z\right)  -c_{\lambda}\left(
u,v\right)  \right)  _{/}%
\]
of the original contra the enforced state-of-opinion $W_{\alpha z}c_{\lambda
}\left(  u,v\right)  $. In Proposition \ref{lim-state} it is proved that when
the enforcement fades away, i.e. when $\alpha\rightarrow0,$ the
state-of-opinion tends to the limit state-of-opinion
\[
\left(  z\#c\left(  u\right)  +\lambda z\#c\left(  v\right)  \right)  _{/}.
\]
By (\ref{dG-.-}) the expected percentage of affirmative answers to $Q$ in this
state is
\begin{align}
&  \mathcal{R}\left(  Q,\left(  z\#c\left(  u\right)  +\lambda z\#c\left(
v\right)  \right)  _{/}\right) \nonumber\\
&  =\frac{\left\langle d\Gamma Q\left(  z\#c\left(  u\right)  +\lambda
z\#c\left(  v\right)  \right)  ,z\#c\left(  u\right)  +\lambda z\#c\left(
v\right)  \right\rangle }{\left\langle d\Gamma I\left(  z\#c\left(  u\right)
+\lambda z\#c\left(  v\right)  \right)  ,z\#c\left(  u\right)  +\lambda
z\#c\left(  v\right)  \right\rangle }\label{Pass-press}\\
&  =\frac{\vartheta^{2}\left(  \lambda,\omega\right)  \left|  Qz\right|
^{2}+\kappa\left(  Q;\lambda,u,v,\omega\right)  -2\lambda\omega\iota\left(
Q,u,v,z\right)  }{\vartheta^{2}\left(  \lambda,\omega\right)  +\kappa\left(
I;\lambda,u,v,\omega\right)  -2\lambda\omega\iota\left(  I,u,v,z\right)
},\nonumber
\end{align}
where $\omega=e^{-\frac{1}{2}\left|  u-v\right|  ^{2}}.$ Due to the lack of
homogeneity relative to $u,v,$ we cannot make $\omega$ in (\ref{Pass-press})
an independent variable as in (\ref{omega-no-omega}).

The term $\iota\left(  Q,u,v,z\right)  $ measures the balance of the influence
of $c\left(  z\right)  $ on components $c\left(  u\right)  $ and $c\left(
v\right)  $ of $c_{\lambda}\left(  u,v\right)  .$ If the influence of
$c\left(  z\right)  $ on $c_{\lambda}\left(  u,v\right)  $ is equally
distributed between $c\left(  u\right)  $ and $c\left(  v\right)  $, the term
$\iota\left(  Q,u,v,z\right)  $ vanishes.

\section{A model\label{model}}

Let us consider a special case. Suppose that in a community two complementary
coherent profiles manifest. Fix two positive numbers $a$ and $b,$ $a^{2}%
+b^{2}=1$ and consider the states-of-mind
\[
u=\mu\binom{a}{b}\text{ and }v=\mu\binom{b}{a}.
\]
Suppose further that the community we investigate is polarized into two
profiles - one in the state $c\left(  u\right)  $ and the other in the state
$c\left(  v\right)  .\ $Let $100a^{2}\%$ of the members of the profile in
state $c\left(  u\right)  $ support an attitude $X$ while $100a^{2}\%$ of the
members of the profile in state $c\left(  v\right)  $ will reject $X$. We
consider the bicoherent state $c_{\lambda}\left(  \mu\binom{a}{b},\mu\binom
{b}{a}\right)  .$

Consider a question corresponding to the projection
\[
Q=\left(
\begin{array}
[c]{cc}%
1 & 0\\
0 & 0
\end{array}
\right)  ,
\]
where the eigenvector $\binom{1}{0}$ implies the answer ''yes'' to the
question
\[
\text{''Do you support the attitude }X\text{?''}%
\]

We shall now analyse the expected relative frequencies of affirmative answers
to $Q$ before and after the exertion of the influence which we choose equally
divided between $c\left(  u\right)  $ and $c\left(  v\right)  $:
\[
z=\frac{1}{\sqrt{2}}\binom{1}{1}.
\]
Hence
\[
\iota\left(  Q,u,v,z\right)  =\iota\left(  I,u,v,z\right)  =0
\]
and
\begin{align*}
\kappa\left(  Q;\lambda,u,v,\omega\right)   &  =\left(  a^{2}+2\lambda\omega
ab+\lambda^{2}b^{2}\right)  \mu^{2}\\
\kappa\left(  I;\lambda,u,v,\omega\right)   &  =\mu^{2}\left(  \left(
1+\lambda^{2}\right)  +4\lambda\omega ab\right)  .
\end{align*}

Using (\ref{R-sub-omega}) we get
\begin{equation}
\mathcal{R}_{\omega}\left(  Q,\lambda,u,v\right)  =\frac{a^{2}+2\lambda\omega
a\sqrt{1-a^{2}}+\lambda^{2}\left(  1-a^{2}\right)  }{\left(  1+\lambda
^{2}\right)  +4\lambda\omega a\sqrt{1-a^{2}}} \label{case 1}%
\end{equation}
and using (\ref{Pass-press}) we get
\begin{align}
&  \mathcal{R}\left(  Q,\left(  z\#c\left(  u\right)  +\lambda z\#c\left(
v\right)  \right)  _{/}\right) \label{case 2}\\
&  \frac{\frac{1}{2}\left(  1+\lambda^{2}+2\lambda\omega\right)  \mu
^{-2}+\left(  a^{2}+2\lambda\omega a\sqrt{1-a^{2}}+\lambda^{2}\left(
1-a^{2}\right)  \right)  }{\left(  1+\lambda^{2}+2\lambda\omega\right)
\mu^{-2}+\left(  \left(  1+\lambda^{2}\right)  +4\lambda\omega a\sqrt{1-a^{2}%
}\right)  },\nonumber
\end{align}
where $\omega=e^{-\frac{1}{2}\left|  u-v\right|  ^{2}}=e^{-\mu^{2}\left(
a-b\right)  ^{2}}$ and $\mu^{2}=\left|  u\right|  ^{2}=\left|  v\right|  ^{2}$
are the energies of $c\left(  u\right)  $ and $c\left(  v\right)  .$

To illustrate the obtained results we shall draw the four pairs of graphs of
(\ref{case 2}) and (\ref{case 1}) imposed on each other, for the choices of
$a^{2}=\frac{19}{36},$ $\frac{22}{36},$ $\frac{26}{36},$ $\frac{28}{36}%
,\ $with respective interaction coefficients $\omega=0.998\,46,$ $0.975\,31,$
$0.901\,05,$ $0.844\,91$. When the energy $\mu$ grows, (\ref{case 2})
converges to (\ref{case 1}) so we can as well choose $\mu=1.$ In each of the
four cases the graph corresponding to (\ref{case 2}) is the one with smaller
maximum and greater minimum than the graph corresponding to (\ref{case
1}).\newline \noindent%
{\includegraphics[
height=1.8343in,
width=2.7389in
]%
{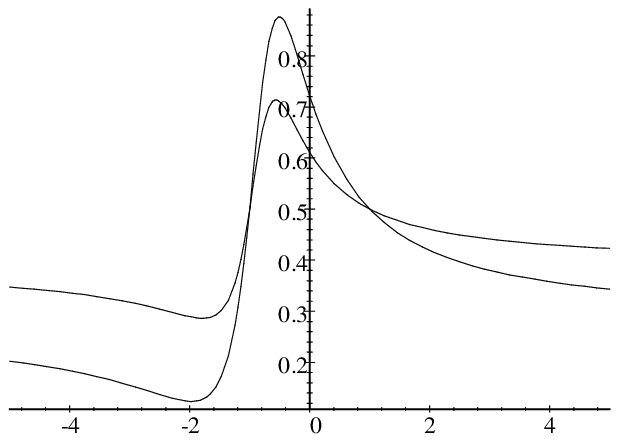}%
}%
{\includegraphics[
height=1.8343in,
width=2.7389in
]%
{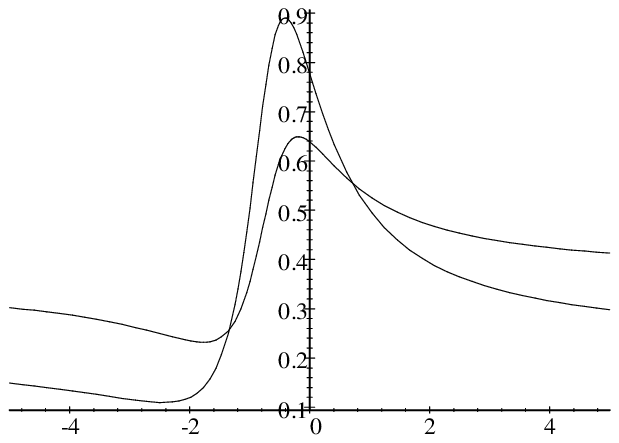}%
}%
\newline
{\includegraphics[
height=1.8343in,
width=2.7389in
]%
{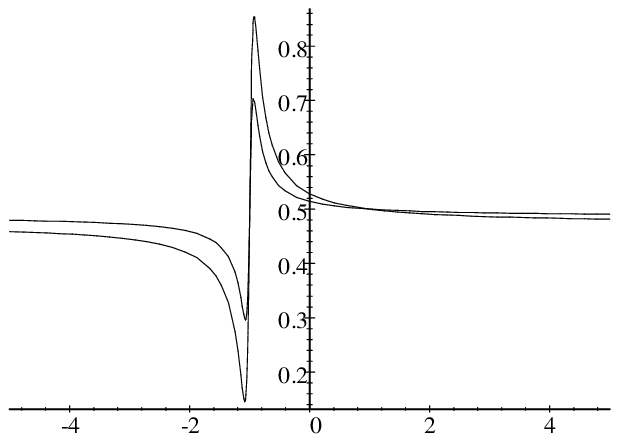}%
}%
{\includegraphics[
height=1.8343in,
width=2.7389in
]%
{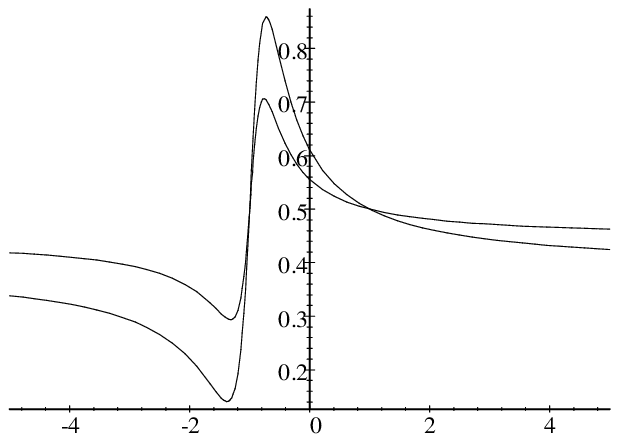}%
}%

Let us look at the diagrams above. For $\lambda>1$ there are no significant
differences in the forms of the diagrams$.$ In all cases maximum is not
attained for $\lambda=0$ but only for a negative $\lambda.$ Movement of
$\lambda$ from zero in the negative direction makes $\mathcal{R}_{\omega}$
increase. Consequently we have

\begin{conclusion}
\textit{The increase of the influence of }$c\left(  v\right)  $\textit{\ acts
as a buster for }$c\left(  u\right)  $\textit{\ providing more affirmative
answers}.
\end{conclusion}

Say the interaction coefficient $\omega$ is close to one. Starting at
$\lambda=0$ and moving in the negative direction, we observe a rapid increase
of $\mathcal{R}_{\omega}$. Then, continuing to move $\lambda$ in the same
direction, the situation reverses - now the maximum decreases towards the minimum.

\begin{conclusion}
\label{switch}\textit{The closer to one is the interaction coefficient, the
shorter is the interval within which }$R_{\omega}$\textit{\ attains first the
maximum and then the minimum.}
\end{conclusion}

Then the situation stabilizes and with further decrease of $\lambda,$
$\mathcal{R}_{\omega}$ converges to its limit $\mathcal{R}\left(  Q,c\left(
v\right)  \right)  $ in $-\infty$ so that the role of $c\left(  u\right)  $ is eliminated.

As an example we can take the government and the opposition of a democratic
country each in a coherent state. Assume that there is an intensive
interaction between $c\left(  u\right)  $ and $c\left(  v\right)  $. Say the
government has majority: $\mathcal{R}\left(  Q,c\left(  u\right)  \right)
>\mathcal{R}\left(  Q,c\left(  v\right)  \right)  .$ A small negative weight
$\lambda$ attached to $c\left(  v\right)  $ yields not much contribution
itself but by way of the interaction it provokes the other fraction to vote.
Similarly if $\mathcal{R}\left(  Q,c\left(  u\right)  \right)  <\mathcal{R}%
\left(  Q,c\left(  v\right)  \right)  ,$ then the increase of negative answers
is provoked. Still higher negative weight yields the reversed status - the
respondents from $c\left(  v\right)  $ take over and in the first case cause a
decrease and in the second case an increase of affirmation. These unusual
variations can happen only in the presence of high interaction and in small
intervals of negative $\lambda$ and hence will seldom occur in real life.
However, such jumps in the distribution of votes have been observed in the
past (cf.\cite{Nannestad}).

It is clearly visible that equally distributed influence of $W_{z}$ makes the
extreme values of $\mathcal{R}_{\omega}$ diminish.

\begin{conclusion}
\label{influence}The influence of $W_{z}$ tempers the voters' reactions.
\end{conclusion}

\bigskip\ 

\label{References}

\end{document}